\documentstyle[epsf,aps]{revtex}
\begin{document}
\begin{flushright}
NUC-MN-99/1-T \\
TPI-MINN-99/05
\end{flushright}
\vspace*{1cm} 
\setcounter{footnote}{1}
\begin{center}
  {\Large\bf Small--$x$ $F_2$ Structure Function of a Nucleus
    Including Multiple Pomeron Exchanges} \\[1cm] Yuri V.\ Kovchegov
  \\ ~~ \\ {\it School of Physics and Astronomy, University of
    Minnesota, \\ Minneapolis, MN 55455 }\\ ~~ \\ ~~ \\ 
\end{center}
\begin{abstract} 
  We derive an equation determining the small-$x$ evolution of the
  $F_2$ structure function of a large nucleus which includes all
  multiple pomeron exchanges in the leading logarithmic approximation
  using Mueller's dipole model \cite{Mueller1,Mueller2,Mueller3,MZ}.
  We show that in the double leading logarithmic limit this evolution
  equation reduces to the GLR equation \cite{GLR,MQ}.
\end{abstract}

\section{Introduction}

The problem of understanding the large gluon density regime in high
energy scattering has always been one of the challenges of
perturbative QCD (PQCD). Unitarity of the total cross--section and
saturation of the gluon distribution are among the most important
issues related to the problem.  The BFKL equation \cite{EAK,Yay} is
the only well--established tool of PQCD which allows us to explore
this high density region by resumming the leading longitudinal
logarithmic contribution to the scattering process. In BFKL evolution
the small--$x$ partons are produced overlapping each other in the
transverse coordinate space \cite{Mupro}, therefore creating high
density regions in the hadron's wave function (hot spots). The
next--to--leading order correction to BFKL equation has been
calculated recently \cite{VSF,Cia}. Although the final conclusion one
should draw from the calculation of \cite{VSF,Cia} is still to be
understood, there are some serious problems associated with the
interpretation of the result \cite{Regge,Levin3,BB}. However, we are
not going to address these issues in this paper for the following
reason. As was shown in \cite{Regge,Levin3} the effects of the second
order BFKL kernel become important in hadron--hadron scattering at the
rapidities of the order of $Y_{NLO} \sim 1 / \alpha^{5/3}$, with
$\alpha$ the strong coupling constant.  At the same time the unitarity
constraints, associated with the multiple (leading order) hard pomeron
exchanges are expected to be reached at $Y_{U} \sim (1 / \alpha) \ln
(1 / \alpha)$ \cite{Mueller5}, which is parametrically smaller than
$Y_{NLO}$ for small coupling constant. Therefore multiple pomeron
exchanges become important at lower center of mass energies than the
effects of subleading corrections, possibly leading to unitarization
of the total hadron--hadron cross--section. Hence the problem of
resummation of the multiple pomeron exchanges seems to be more
important for describing the recent experimental results, such as ZEUS
1995 data \cite{Cald}, which probably shows evidence of saturation of
the $F_2$ structure function at low $Q^2$.

In this paper we are going to consider deep inelastic scattering (DIS)
of a virtual photon on a large nucleus and will resum all multiple
pomeron exchanges contributing to the $F_2$ structure function of the
nucleus in the leading longitudinal logarithmic approximation in the
large $N_c$ limit. The first step in that direction in PQCD was the
derivation by Gribov, Levin and Ryskin (GLR) of an equation describing
the fusion of two pomeron ladders into one in the double logarithmic
approximation \cite{GLR}.  The resulting equation with the coefficient
in front of the quadratic term calculated by Mueller and Qiu \cite{MQ}
for a low density picture of a spherical proton of radius $R$ reads
\begin{equation}\label{GLReq}
  \frac{\partial^2 xG (x, Q^2) }{\partial \ln (1/x)\, \partial \ln
    (Q^2 / \Lambda^2_{QCD}) } = \frac{\alpha N_c}{\pi} xG (x, Q^2) -
  \frac{4 \alpha^2 N_c}{3 C_F R^2} \frac{1}{Q^2} [xG (x, Q^2)]^2.
\end{equation}
This equation sums up all multiple hard pomeron exchanges in the
gluon distribution function in the double logarithmic limit.

\begin{figure}
\begin{center}
\epsfysize=5cm
\leavevmode
\hbox{ \epsffile{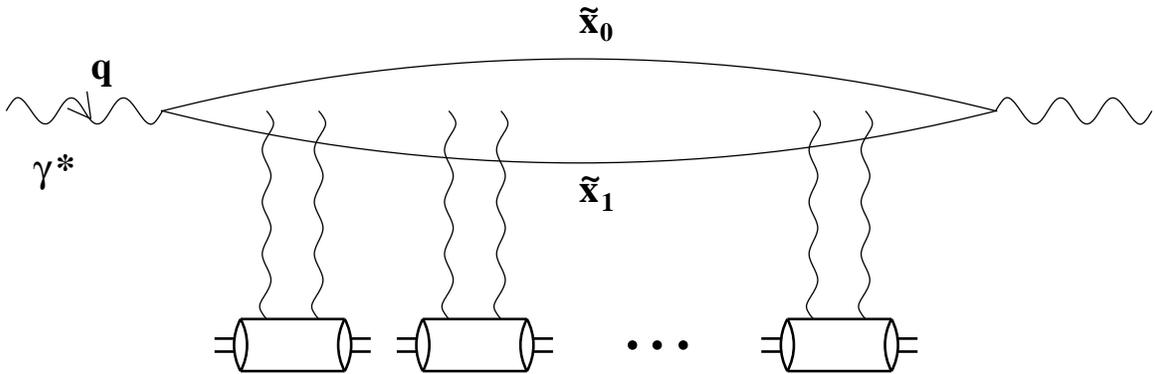}}
\end{center}
\caption{Forward amplitude of DIS on a nucleus in the 
  quasi--classical (Glauber) approximation.}
\label{glauber}
\end{figure}

Since then there have been several attempts to generalize the GLR
equation. Recently an equation has been proposed by Ayala, Gay Ducati,
and Levin in \cite{Levin1,Levin2}, which tries to incorporate the
Glauber--type multiple rescatterings of a probe on the nucleons in a
nucleus (see Fig. \ref{glauber}). Using the results of Mueller in
\cite{Mueller4} for a pair of gluons multiply rescattering inside a
nucleus, the authors of \cite{Levin1,Levin2} obtained the following
equation for the gluon distribution of the nucleus in the double
logarithmic approximation
\begin{eqnarray}\label{glamu1}
  \frac{\partial^2 x G_A (x, Q^2)}{\partial \ln (Q^2 / \Lambda_{QCD}^2
    ) \partial \ln (1/x) } = \frac{N_c C_F S_\perp}{\pi^3} Q^2 \left\{
    1 - \exp \left[ - \frac{2 \alpha \pi^2}{N_c S_\perp} \frac{1}{Q^2}
      xG_A ( x , Q^2 ) \right] \right\}.
\end{eqnarray}
If one expands the right hand side of Eq. (\ref{glamu1}) to the second
order in $xG_A$ one recovers the GLR equation [Eq. (\ref{glamu1}) is
written here for a cylindrical nucleus. Therefore the coefficients in the
obtained GLR equation will not match those of Eq. (\ref{GLReq})].

An extensive work on resumming the multiple pomeron exchanges in the
gluon distribution function in the leading $\ln (1/x)$ approximation
(i.e. without taking the double logarithmic limit) has been pursued by
Jalilian-Marian, Kovner, Leonidov, McLerran, Venugopalan, and Weigert
\cite{MV,Larry1,JKLW1,JKLW2,JKW,JKLW}. Starting with a model of a
large nucleus \cite{MV,me}, which provides some effective action
\cite{MV,Larry1}, they develop a renormalization group procedure which
integrates out harder longitudinal gluonic degrees of freedom in the
nucleus and allows one to resum leading $\ln (1/x)$ contribution to
the gluon distribution function. The resulting equation, written in a
functional form in \cite{JKLW2} is supposed to resum these leading
logarithms including also all powers of color charge density of the
nucleus, which in a more traditional language corresponds to
resummation of multiple pomeron exchanges. However, even though that
equation at the lowest (one pomeron) level reduces to the expected
BFKL equation \cite{JKLW1}, in general it is a very complicated
functional differential equation which can not be solved even
numerically. Recently the double logarithmic limit of that equation
was obtained \cite{JKLW}, providing us with another equation for $xG$:
\begin{eqnarray*}
  \frac{\partial^2 x G (x, Q^2)}{\partial \ln (Q^2 / \Lambda_{QCD}^2 )
    \partial \ln (1/x) } = \frac{N_c (N_c - 1)}{2} S_\perp Q^2 \left[
    1 + \frac{\pi (N_c - 1) Q^2 S_\perp}{2 \alpha xG} \exp \left(
      \frac{\pi (N_c - 1) Q^2 S_\perp}{2 \alpha xG} \right) \right.
\end{eqnarray*}
\begin{eqnarray}\label{alik}
  \left. \times \mbox{Ei}\left( - \frac{\pi (N_c - 1) Q^2 S_\perp}{2
        \alpha xG} \right) \right].
\end{eqnarray}
Eq. (\ref{alik}) is written here for a cylindrical nucleon with
transverse area $S_\perp$. In the limit of small gluon density Eq.
(\ref{alik}) reduces to GLR equation \cite{JKLW}.

Our approach in this paper is pursuing the same goals as authors of
\cite{MV,Larry1,JKLW1,JKLW2,JKW,JKLW}. We will also write an equation
which resums all multiple pomeron exchanges on a nucleus in the
leading logarithmic approximation. However our strategy is a bit
different from \cite{MV,Larry1,JKLW1,JKLW2,JKW,JKLW}. We will consider
the scattering of a virtual photon on a nucleus at rest, therefore
putting all the QCD evolution in the wave function of the virtual
photon. This is different from what was done by the authors of
\cite{MV,Larry1,JKLW1,JKLW2,JKW,JKLW}, since they were developing the
QCD evolution inside the nucleus. The virtual photon's wave function
including the leading logarithmic evolution was constructed in the
large $N_c$ limit by Mueller in \cite{Mueller1,Mueller2,Mueller3,MZ}.
This so--called dipole wave function in fact contains all multiple
pomeron exchanges, which in the large $N_c$ language can be pictured
as multiple color ``cylinders''. A numerical analysis of the
unitarization of the total onium--onium cross--section through
multiple pomeron exchanges was carried out in the framework of the
dipole model by Salam in \cite{Salam}. Considering the scattering of a
virtual photon (quarkonium state) on a nucleus simplifies the problem,
allowing us to treat it analytically.  In Sect. II we will use the
dipole wave function to write down an equation which governs the
evolution of the $F_2$ structure function of the nucleus (formula
(\ref{eqN})). Our equation is directly related to a physical
observable ($F_2$) and, therefore, is free from all the problems and
ambiguities associated with dealing with the gluon distribution
function $xG$. It is a non--linear integral equation, not a functional
differential equation like in \cite{JKLW1}. Therefore one should be
able to solve Eq.  (\ref{eqN}) at least numerically.

We will dedicate Sect. III to exploring the equation resulting from
taking double logarithmic (large $Q^2$) limit of the equation derived
in Sect. II. We see that our equation reduces to GLR equation, failing
to reproduce Eqs. (\ref{glamu1}), (\ref{alik}). Finally, in Sect. IV
we will conclude by discussing the limitations of the large $N_c$
approximation, as well as some advantages of our approach.

\section{Evolution equation from the dipole model}

We start by considering a deep inelastic scattering process on a
nucleus. As shown in Fig.\ref{glauber} , the incoming virtual photon
with a large $q_+$ component of the momentum splits into a
quark--antiquark pair which then interacts with the nucleus at rest.
We model the interaction by no more than two gluon exchanges between
each of the nucleons and the quark--antiquark pair. This is done in
the spirit of the quasi--classical approximation used previously in
\cite{Mueller4,mM,BDMPS}. The interactions are taken in the eikonal
approximation. Then, as could be shown in general, i.e., including the
leading logarithmic QCD evolution, the total cross--section, and,
therefore, the $F_2$ structure function of the nucleus can be
rewritten as a product of the square of the virtual photon's wave
function and the propagator of the quark--antiquark pair through the
nucleus \cite{KKMV,NZ}.The expression reads \cite{KKMV}
\begin{equation}\label{defN}
  F_2 (x, Q^2) = \frac{Q^2}{4 \pi^2 \alpha_{EM}} \int \frac{d^2 {\bf
      x}_{01} d z }{2 \pi} \, [\Phi_T ({\bf x}_{01},z) + \Phi_L ({\bf
    x}_{01},z) ] \ d^2 b_0 \ N({\bf x}_{01},{\bf b}_0 , Y) ,
\end{equation}
where the incoming photon with virtuality $Q$ splits into a
quark--antiquark pair with the transverse coordinates of the quark and
antiquark being ${\bf \tilde{x}}_0$ and ${\bf \tilde{x}}_1$
correspondingly, such that ${\bf x}_{10} = {\bf \tilde{x}}_1 - {\bf
  \tilde{x}}_0$. The coordinate of the center of the pair is given by
${\bf b}_0 = \frac{1}{2} ( {\bf\tilde{x} }_1 + {\bf \tilde{x}}_0 )$.
$Y$ is the rapidity variable $Y = \ln s/Q^2 = \ln 1/x$. The square of
the light cone wave function of $q \overline{q}$ fluctuations of a
virtual photon is denoted by $\Phi_T ({\bf x}_{01},z)$ and $\Phi_L
({\bf x}_{01},z)$ for transverse and longitudinal photons
correspondingly, with $z$ being the fraction of the photon's
longitudinal momentum carried by the quark. At the lowest order in
electromagnetic coupling ($\alpha_{EM}$) $\Phi_T ({\bf x}_{01},z)$ and
$\Phi_L ({\bf x}_{01},z)$ are given by [\cite{KKMV,NZ} and references
therein]
\begin{mathletters}\label{wfn}
\begin{equation}
  \Phi_T ({\bf x}_{01},z) = \frac{2 N_c \alpha_{EM}}{\pi} \left\{ a^2
    \ K_1^2 (x_{01} a) \ [z^2 + (1 - z)^2] \right\},
\end{equation}
\begin{equation}
  \Phi_L ({\bf x}_{01},z) = \frac{2 N_c \alpha_{EM}}{\pi} \ 4 Q^2 z^2
  (1 - z)^2 \ K_0^2 (x_{01} a),
\end{equation}
\end{mathletters}
with $a^2 = Q^2 z (1 - z)$. We consider massless quarks having only
one flavor.

The quantity $N({\bf x}_{01},{\bf b}_0, Y)$ has the meaning of the
forward scattering amplitude of the quark--antiquark pair on a nucleus
\cite{NZ}. At the lowest (classical) order not including the QCD
evolution in rapidity it is given by
\begin{mathletters}\label{gamma}
\begin{equation}\label{gamma1}
  N({\bf x}_{01},{\bf b}_0, 0) = - \gamma ({\bf x}_{01},{\bf b}_0)
  \equiv \left\{ 1 - \exp \left[ - \frac{C_F}{N_c} \frac{{\bf
          x}_{01}^2 \tilde{v} (x_{01}) R}{2 \lambda}\right] \right\},
\end{equation}
with $\tilde{v}$ as defined in \cite{mM} and $\lambda$ being the mean
free path of a gluon in a nuclear medium, as defined in \cite{mM}. In
the logarithmic approximation for large $Q^2$ (small $x_{01}$) Eq.
(\ref{gamma1}) can be rewritten as
\begin{equation}
  N({\bf x}_{01},{\bf b}_0, 0) = - \gamma ({\bf x}_{01},{\bf b}_0)
  \approx \left\{ 1 - \exp \left[ - \frac{\alpha \pi^2}{2 N_c S_\perp}
      {\bf x}_{01}^2 A xG ( x ,1/{\bf x}_{01}^2 ) \right] \right\}.
\end{equation}
\end{mathletters}
$\gamma ({\bf x}_{01},{\bf b}_0)$ is the propagator of the $q
\overline{q}$ pair through the nucleus. The propagator could be easily
calculated, similarly to \cite{mM,KKMV}, giving the Glauber multiple
rescattering formula (\ref{gamma}). Here and throughout the paper we
assume for simplicity that the nucleus is a cylinder, which appears as
a circle of radius $R$ in the transverse direction and has a constant
length $2 R$ along the longitudinal $z$ direction.  Therefore its
transverse cross--sectional area is $S_\perp = \pi R^2$.  In formula
(\ref{gamma}) $A$ is the atomic number of the nucleus, $\alpha$ is the
strong coupling constant and $xG ( x ,1/{\bf x}_{01}^2 )$ is the gluon
distribution in a nucleon in the nucleus, taken at the lowest order in
$\alpha$, similarly to \cite{mM}.

Eq. (\ref{gamma}) resums all Glauber type multiple rescatterings of a
$q \overline{q}$ pair on a nucleus. As was mentioned before, since
each interaction of the pair with a nucleon in the nucleus is
restricted to the two gluon exchange, the formula (\ref{gamma})
effectively sums up all the powers of the parameter $\alpha^2
A^{1/3}$. Or, looking at the power of the exponent in (\ref{gamma}) we
conclude that since $x_{01} \sim 1/Q$, it resums all the powers of
$\frac{\alpha^2 A^{1/3}}{Q^2}$. This is the definition of
quasi--classical limit, a more detailed discussion of which could be
found in \cite{me}.

\begin{figure}
\begin{center}
\epsfysize=7cm
\leavevmode
\hbox{ \epsffile{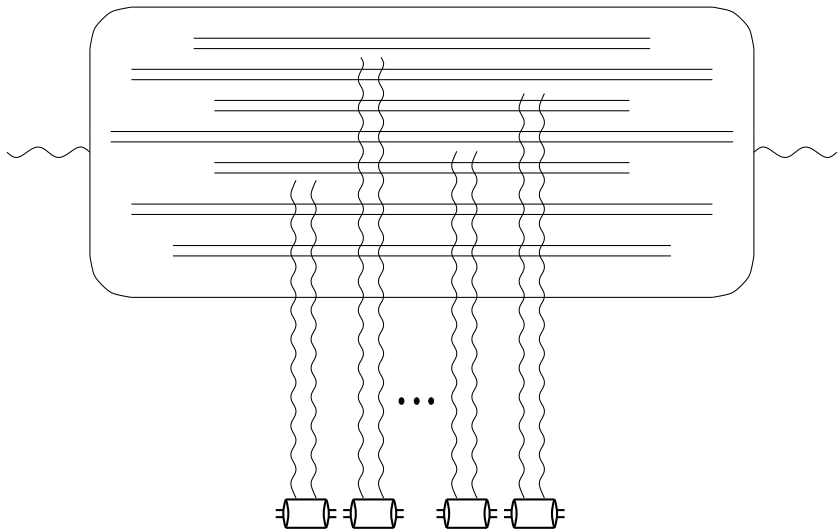}}
\end{center}
\caption{DIS on a nucleus including the QCD evolution in the
quark--antiquark pair in the form of dipole wave function. Each 
double line represents a gluon in the large--$N_c$ limit.}
\label{dipole}
\end{figure}

Since the nucleus is at rest in order to include the QCD evolution of
$F_2$ structure function, we have to develop the soft gluon wave
function of the incoming virtual photon. In the leading longitudinal
logarithmic approximation ($\ln 1/x$) the evolution of the wave
function is realized through successive emissions of small--$x$
gluons.  The $q \overline{q}$ pair develops a cascade of gluons, which
then scatter on the nucleus. In order to describe the soft gluon
cascade we will take the limit of a large number of colors, $N_c
\rightarrow \infty$. Then, this leading logarithmic soft gluon
wavefunction will become equivalent to the dipole wave function,
introduced by Mueller in \cite{Mueller1,Mueller2,Mueller3}. The
physical picture becomes straightforward. The $q \overline{q}$ pair
develops a system of dipoles (dipole wave function), and each of the
dipoles independently scatters on the nucleus, as shown in Fig.
\ref{dipole}. Since the nucleus is large, we may approximate the
interaction of a dipole (quark--antiquark pair) with the nucleus by
$\gamma ({\bf x},{\bf b})$ given by Eq.  (\ref{gamma}), with ${\bf x}$
and ${\bf b}$ being the dipole's transverse separation and impact
parameter. That means that each of the dipoles interacts with several
nucleons (Glauber rescattering) in the nucleus independent of other
dipoles. The interaction of each of the dipoles with the nucleus is
the same as was shown in Fig. \ref{glauber} for the initial $q
\overline{q}$ pair.

To construct the dipole wave function we will heavily rely on the
techniques developed in \cite{Mueller1,Mueller2,Mueller3,MZ}.
Following \cite{Mueller1,Mueller3} we define the generating functional
for dipoles $Z ({\bf b}_0,{\bf x}_{01}, Y, u)$ (see formulae (16) and
(17) in \cite{Mueller1}). The generating functional then obeys the
equation (see Eq.  (12) in \cite{Mueller3})
\begin{eqnarray*}
  Z ({\bf b}_0,{\bf x}_{01}, Y, u) = u ({\bf b}_0,{\bf x}_{01} ) \exp
  \left[ - \frac{4 \alpha C_F}{\pi} \ln \left( \frac{x_{01}}{\rho}
    \right) Y \right] +
\end{eqnarray*}
\begin{eqnarray*}
  + \frac{\alpha C_F}{\pi^2} \int_0^Y d y \exp \left[ - \frac{4 \alpha
      C_F}{\pi} \ln \left( \frac{x_{01}}{\rho} \right) (Y - y) \right]
  \end{eqnarray*}
\begin{eqnarray}\label{ZZ}
  \times \int_\rho d^2 \tilde{x}_2 \frac{x_{01}^2}{x_{02}^2 x_{12}^2}
  \, Z ({\bf b}_0 + \frac{1}{2}{\bf x}_{12} ,{\bf x}_{02}, y, u) \, Z
  ({\bf b}_0 -\frac{1}{2}{\bf x}_{20} ,{\bf x}_{12}, y, u),
\end{eqnarray}
where ${\bf x}_{20} = {\bf \tilde{x}}_0 - {\bf \tilde{x}}_2$ , ${\bf
  x}_{21}= {\bf \tilde{x}}_1 - {\bf \tilde{x}}_2$ and the integration
over $\tilde{x}_2$ is performed over the region where $x_{02} \geq
\rho$ and $x_{12} \geq \rho$. This $\rho$ serves as an ultraviolet
cutoff in the equation and disappears in the physical quantities.
${\bf b}_0 = \frac{1}{2} ({\bf x}_0 +{\bf x}_1 )$ is the position of
the center of the initial dipole in the transverse plane
\cite{Mueller3}. $C_F = N_c / 2$ in the large $N_c$ limit. The
generating functional is defined such that $Z ({\bf b}_0,{\bf x}_{01},
Y, u = 1) = 1$ [see \cite{Mueller1}].

Analogous to \cite{Mueller2,Mueller3} we now define the dipole number
density by
\begin{equation}\label{n1}
  \frac{1}{2 \pi x^2} n_1 (x_{01}, Y, |{\bf b} - {\bf b}_0|, x ) =
  \frac{\delta}{\delta u ({\bf b}, {\bf x})} Z ({\bf b}_0,{\bf
    x}_{01}, Y, u) |_{u = 1}.
\end{equation}
$n_1 (x_{01}, Y, |{\bf b} - {\bf b}_0|, x )$ convoluted with the virtual
photon's wave function gives the number of dipoles of transverse size
$x$ at the impact parameter $|{\bf b} - {\bf b}_0|$ with the smallest
light cone momentum in the pair greater or equal to $e^{-Y} q_+$.
Similarly to the dipole number density we can introduce dipole pair
density \cite{Mueller2,Mueller3} for a pair of dipoles of sizes $x_1$
and $x_2$ at the impact parameters $|{\bf b}_1 - {\bf b}_0|$ and$|{\bf
  b}_2 - {\bf b}_0|$ by
\begin{equation}\label{dippair}
  \frac{1}{2 \pi x_1^2} \frac{1}{2 \pi x_2^2} n_2 (x_{01}, Y, |{\bf
    b}_1 - {\bf b}_0|, x_1 , |{\bf b}_2 - {\bf b}_0|, x_2 ) =
  \frac{1}{2!} \frac{\delta}{\delta u ({\bf b}_1, {\bf x}_1)}
  \frac{\delta}{\delta u ({\bf b}_2, {\bf x}_2)} Z ({\bf b}_0,{\bf
    x}_{01}, Y, u) |_{u = 1}.
\end{equation}
Our notation is different from the conventional approach of
\cite{Mueller2,Mueller3} by the factor of a factorial, for reasons
which will become obvious later. Generalizing the definition
(\ref{dippair}) to $k$ dipoles of sizes $x_1, \ldots , x_n$ situated
at the impact parameters $|{\bf b}_1 - {\bf b}_0|, \ldots , |{\bf b}_k
- {\bf b}_0|$ we easily obtain:
\begin{equation}\label{kdip}
  \prod_{i = 1}^k \frac{1}{2 \pi x_i^2} n_k (x_{01}, Y, |{\bf b}_1 -
  {\bf b}_0|, x_1 , \ldots , |{\bf b}_k - {\bf b}_0|, x_k ) =
  \frac{1}{k!} \prod_{i = 1}^k \frac{\delta}{\delta u ({\bf b}_i,
    {\bf x}_i)} \, Z ({\bf b}_0,{\bf x}_{01}, Y, u) |_{u = 1}.
\end{equation}

One can now see that in order to include all the multiple pomeron
exchanges one has to sum up the contributions of different numbers of
dipoles interacting with the nucleus. Namely we should take the dipole
number density $n_1 (x_{01}, Y, {\bf b}, x)$ and convolute it with the
propagator of this one dipole in the nucleus $\gamma ( {\bf x} ,{\bf
  b})$.  Then we should take the dipole pair density $n_2(x_{01}, Y,
{\bf b}_1 , x_1 , {\bf b}_2 , x_2 )$ and convolute it with two
propagators $\gamma ( {\bf x}_1 ,{\bf b}_1 )$ and $\gamma ( {\bf x}_2
,{\bf b}_2 )$, etc. That way we obtain an expression for $N({\bf
  x}_{01},{\bf b}_0, Y)$:
\begin{eqnarray*}
  - \, N({\bf x}_{01},{\bf b}_0, Y) = \int n_1 (x_{01}, Y, {\bf b}_1,
  {\bf x}_1) \left( \gamma ( {\bf x}_1 ,{\bf b}_1) \frac{d^2 x_1}{2
      \pi x_1^2} d^2 b_1 \right) +
\end{eqnarray*}
\begin{eqnarray*}
  + \int n_2(x_{01}, Y, {\bf b}_1 , {\bf x}_1 , {\bf b}_2 , {\bf x}_2
  ) \left( \gamma ( {\bf x}_1 ,{\bf b}_1) \frac{d^2 x_1}{2 \pi x_1^2}
    d^2 b_1 \right) \left( \gamma ( {\bf x}_2 ,{\bf b}_2) \frac{d^2
      x_2}{2 \pi x_2^2} d^2 b_2 \right) + \ldots = 
\end{eqnarray*}
\begin{eqnarray}\label{NN}
  = \sum_{i=1}^\infty \int n_i (x_{01}, Y, {\bf b}_1 , {\bf x}_1 ,
  \ldots , {\bf b}_i , {\bf x}_i ) \left( \gamma ( {\bf x}_1 ,{\bf
      b}_1) \frac{d^2 x_1}{2 \pi x_1^2} d^2 b_1 \right) \ldots \left(
    \gamma ( {\bf x}_i ,{\bf b}_i) \frac{d^2 x_i}{2 \pi x_i^2} d^2 b_i
  \right),
\end{eqnarray}
where we put the minus sign in front of $N$ to make it positive, since
$\gamma$ is negative. Eq. (\ref{NN}) clarifies the physical meaning of
$N$ as a total cross--section of a $q \overline{q}$ pair interacting
with a nucleus.  One can understand now the factorials in the
definitions of the dipole number densities (\ref{n1}), (\ref{dippair})
and (\ref{kdip}): once the convolutions with the propagators $\gamma$
are done then the dipoles become ``identical'' and we have to include
the symmetry factors.

In order to write down an equation for $N({\bf x}_{01},{\bf b}_0, Y)$
we have to find the equations for $n_i$'s first. Following the
techniques introduced in \cite{Mueller1,Mueller2,Mueller3,MZ} we have
to differentiate the equation for the generating functional (\ref{ZZ})
with respect to $u ({\bf x},{\bf b})$ putting $u = 1$ at the end,
keeping in mind that $Z ({\bf b}_0,{\bf x}_{01}, Y, u = 1) = 1$.
Differentiating formula (\ref{ZZ}) once we obtain an equation for $n_1
(x_{01}, Y, {\bf b}_1, {\bf x}_1)$:
\begin{eqnarray*}
  n_1 (x_{01}, Y, {\bf b}_1, {\bf x}_1) = \delta^2 ({\bf x}_{01} -{\bf
    x}_1) \, 2 \pi {\bf x}_1^2 \, \delta^2 ({\bf b}_1) \exp \left[ -
    \frac{4 \alpha C_F}{\pi} \ln \left( \frac{x_{01}}{\rho} \right) Y
  \right] +
\end{eqnarray*}
\begin{eqnarray}\label{eqn1}
  + \frac{\alpha C_F}{\pi^2} \int_0^Y d y \, \exp \left[ - \frac{4
      \alpha C_F}{\pi} \ln \left( \frac{x_{01}}{\rho} \right) (Y - y)
  \right] \int_\rho d^2 \tilde{x}_2 \frac{x_{01}^2}{x_{02}^2 x_{12}^2}
  \, 2 \, n_1 (x_{02}, y, {\overline {\bf b}}_1, {\bf x}_1) ,
\end{eqnarray}
where, following \cite{Mueller3}, we have defined ${\overline {\bf
    b}}_i = {\bf b}_i -{\bf b}_0 - \frac{1}{2} {\bf x}_{12}$.

Differentiating Eq. (\ref{ZZ}) twice we obtain an equation for
$n_2(x_{01}, Y, {\bf b}_1 , {\bf x}_1 , {\bf b}_2 , {\bf x}_2 )$
\begin{eqnarray*}
  n_2(x_{01}, Y, {\bf b}_1 , {\bf x}_1 , {\bf b}_2 , {\bf x}_2 ) =
  \frac{\alpha C_F}{\pi^2} \int_0^Y d y \, \exp \left[ - \frac{4
      \alpha C_F}{\pi} \ln \left( \frac{x_{01}}{\rho} \right) (Y - y)
  \right] \int_\rho d^2 \tilde{x}_2 \frac{x_{01}^2}{x_{02}^2 x_{12}^2}
\end{eqnarray*}  
\begin{eqnarray}\label{eqn2}
  \times [ 2 \, n_2(x_{02}, y, {\overline {\bf b}}_1 , {\bf x}_1 ,
  {\overline {\bf b}}_2, {\bf x}_2 ) + n_1 (x_{02}, y, {\overline {\bf
      b}}_1 , {\bf x}_1) \, n_1 (x_{12}, y, {\tilde {\bf b}}_2 , {\bf
    x}_2) ],
\end{eqnarray}
where ${\tilde {\bf b}}_i = {\bf b}_i -{\bf b}_0 + \frac{1}{2} {\bf
  x}_{20}$. Now higher order differentiation of Eq.  (\ref{ZZ})
becomes apparent, and could be easily done yielding the following
equation for the number density of $i$ dipoles:
\begin{eqnarray*}
  n_i(x_{01}, Y, {\bf b}_1 , {\bf x}_1 , \ldots , {\bf b}_i , {\bf x}_i ) =
  \frac{\alpha C_F}{\pi^2} \int_0^Y d y \, \exp \left[ - \frac{4
      \alpha C_F}{\pi} \ln \left( \frac{x_{01}}{\rho} \right) (Y - y)
  \right] \int_\rho d^2 \tilde{x}_2 \frac{x_{01}^2}{x_{02}^2 x_{12}^2}
\end{eqnarray*}  
\begin{eqnarray}\label{eqni}
  \times [ 2 \, n_i(x_{02}, y, {\overline {\bf b}}_1 , {\bf x}_1 ,
  \ldots , {\overline {\bf b}}_i , {\bf x}_i ) + \sum_{j + k =i} n_j
  (x_{02}, y, {\overline {\bf b}}_1, {\bf x}_1 , \ldots , {\overline
    {\bf b}}_j , {\bf x}_j ) \, n_k (x_{12}, y, {\tilde {\bf
      b}}_{j+1}, {\bf x}_{j+1}, \ldots , {\tilde {\bf b}}_i , {\bf
    x}_i ) ] ,
\end{eqnarray}
where we anticipate the integration over the dipole sizes and treat
the dipoles as identical objects. In principle Eq. (\ref{eqni}) should
contain the permutations of the arguments of the gluon densities in
the product on the right hand side, but for the abovementioned reason
we do not write this terms explicitly.

   Multiplying formula (\ref{eqni}) by 
\begin{eqnarray*}
  \left( \gamma ( {\bf x}_1 ,{\bf b}_1) \frac{d^2 x_1}{2 \pi x_1^2}
    d^2 b_1 \right) \ldots \left( \gamma ( {\bf x}_i ,{\bf b}_i)
    \frac{d^2 x_i}{2 \pi x_i^2} d^2 b_i \right),
\end{eqnarray*}
integrating over the dipole sizes and impact parameters, and summing
all such equations, i.e. summing over $i$ from 1 to $\infty$ in
(\ref{eqni}) one obtains the equation for $N({\bf x}_{01},{\bf b}_0,
Y)$
\begin{eqnarray*}
  N({\bf x}_{01},{\bf b}_0, Y) = - \gamma ({\bf x}_{01},{\bf b}_0) \,
  \exp \left[ - \frac{4 \alpha C_F}{\pi} \ln \left(
      \frac{x_{01}}{\rho} \right) Y \right] + \frac{\alpha C_F}{\pi^2}
  \int_0^Y d y \, \exp \left[ - \frac{4 \alpha C_F}{\pi} \ln \left(
      \frac{x_{01}}{\rho} \right) (Y - y) \right]
\end{eqnarray*}  
\begin{eqnarray}\label{eqN}
  \times \int_\rho d^2 \tilde{x}_2 \frac{x_{01}^2}{x_{02}^2 x_{12}^2}
  \, [ 2 \, N({\bf x}_{02},{\bf b}_0 + \frac{1}{2} {\bf x}_{12}, y) -
  N({\bf x}_{02},{\bf b}_0 + \frac{1}{2} {\bf x}_{12}, y) \, N({\bf
    x}_{12},{\bf b}_0 - \frac{1}{2} {\bf x}_{20}, y) ] .
\end{eqnarray}
Equation (\ref{eqN}), together with equations (\ref{defN}) and
(\ref{wfn}), provide us with the leading logarithmic evolution of the
$F_2$ structure function of a nucleus including all multiple pomeron
exchanges in the large--$N_c$ limit.

\begin{figure}
\begin{center}
\epsfxsize=12cm
\leavevmode
\hbox{ \epsffile{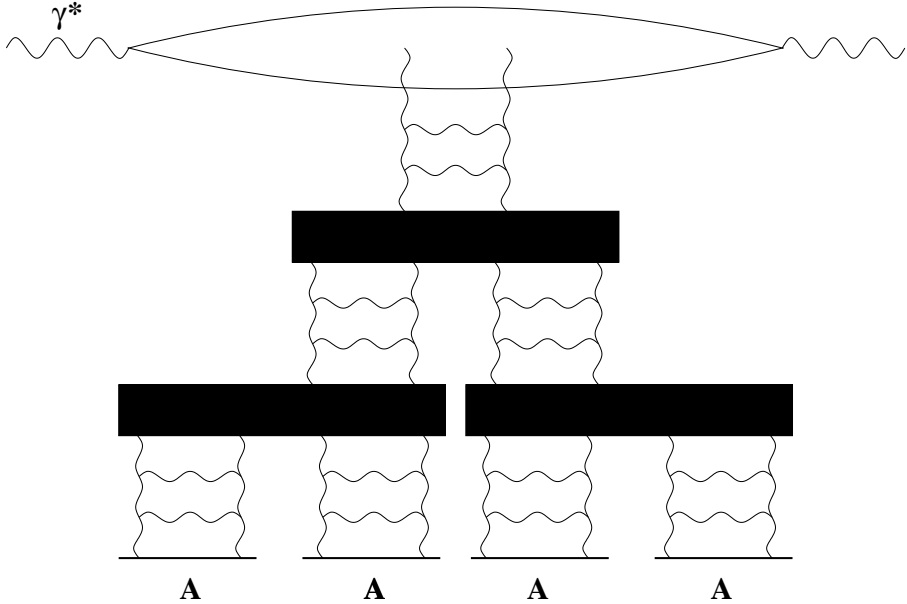}}
\end{center}
\caption{Multiple pomeron exchanges and splittings resummed by 
Eq. (\ref{eqN}). Each pomeron ladder interacts with a nucleus, 
which is symbolically denoted by $A$.}
\label{fan}
\end{figure}

Throughout the preceding calculations we never made an assumption that
$Q^2$ is large. Of course it should be large enough for the
perturbation theory to be applicable.  The only assumption about the
incoming photon's momentum that we made was that its light--cone
component $q_+$ is large, therefore we could neglect the inverse
powers of $q_+$. This is eikonal approximation, which is natural for
leading $\ln (1/x)$ calculation. However if the inverse power of $q_+$
comes with an inverse power of $q_-$, forming something like $1 / 2
q_+ q_- \sim 1 / Q^2 $ we do not neglect these terms, therefore
resumming all the inverse powers of $Q^2$ (``higher twist terms'').
That way we proceed to conclude that equation (\ref{eqN}) sums up in
the leading logarithmic approximation all diagrams that include the
effects of multiple pomeron exchanges, with pomeron ladders together
with pomeron splitting vertices being incorporated in the dipole wave
function. In terms of conventional (not ``wave functional'') language
Eq.  (\ref{eqN}) resums the so called ``fan'' diagrams (see Fig.
\ref{fan}) which were summed up by conventional GLR equation
\cite{GLR,MQ}. The difference between our equation and GLR is that Eq.
(\ref{eqN}) does not assume leading transverse logarithmic (large
$Q^2$) approximation.

\section{Double logarithmic limit}

In order to reconcile our approach with traditional results in this
section we will take the large $Q^2$ limit of Eq. (\ref{eqN}) and show
that in this double logarithmic approximation Eq. (\ref{eqN}) reduces
to the GLR equation \cite{GLR,MQ}. We consider a scattering of a
virtual photon, characterized by large momentum scale $Q$, on a
nucleus at rest characterized by the scale $\Lambda_{QCD}$. The $Q^2
\gg \Lambda_{QCD}^2$ limit implies that the dipoles produced at each
step of the evolution in the dipole wave function must be of much
greater transverse dimensions than the dipoles off which they were
produced. Basically, since in the double logarithmic approximation the
transverse momentum of the gluons in the dipole wave function should
evolve from the large scale $Q$ to the small scale $\Lambda_{QCD}$,
than the transverse sizes of the dipoles should evolve from the small
scale $1/Q$ to the large scale $1/\Lambda_{QCD}$.

In the limit when the produced dipoles are much larger than the dipole
by which they were produced (large $Q^2$ limit), the kernel of Eq.
(\ref{eqN}) becomes
\begin{eqnarray}\label{lda}
  \int_\rho d^2 \tilde{x}_2 \frac{x_{01}^2}{x_{02}^2 x_{12}^2}
  \rightarrow x_{01}^2 \pi \int_{x_{01}^2}^{1/\Lambda^2_{QCD}} \frac{d
    x_{02}^2}{(x_{02}^2)^2},
\end{eqnarray}
where $x_{02} \approx x_{12} \gg x_{01}$, and the upper cutoff of the
$x_{02}$ integration is given by the inverse momentum scale
characterizing the nucleus, $1/\Lambda^2_{QCD}$. Since this
integration is done in the region of large transverse sizes the
ultraviolet cutoff $\rho$ is no longer needed. One can easily see that
including virtual corrections would bring in the exponential factor
$e^{- \frac{\alpha C_F}{\pi}Y}$ in the Eqs. (\ref{ZZ}) and (\ref{eqN})
instead of $\exp \left[ - \frac{4 \alpha C_F}{\pi} \ln \left(
    \frac{x_{01}}{\rho} \right) Y \right]$. In the double logarithmic
approximation $\alpha Y \ln (Q^2/\Lambda^2_{QCD}) \ge 1$ and $\ln
(Q^2/\Lambda^2_{QCD}) \gg 1$, therefore $\alpha Y \le 1$. That way the
factor of $e^{- \frac{\alpha C_F}{\pi}Y}$ can be neglected.  The
resulting limit of Eq.  (\ref{eqN}) is
\begin{eqnarray*}
  N({\bf x}_{01},{\bf b}_0, Y) = - \gamma ({\bf x}_{01},{\bf b}_0) \,
  + \frac{\alpha C_F}{\pi} x_{01}^2 \int_0^Y d y \,
  \int_{x_{01}^2}^{1/\Lambda^2_{QCD}} \, \frac{d
    x_{02}^2}{(x_{02}^2)^2} \, [ 2 \, N({\bf x}_{02},{\bf b}_0, Y)
\end{eqnarray*}
\begin{eqnarray*}
  - N({\bf x}_{02},{\bf b}_0, Y) \, N({\bf x}_{02},{\bf b}_0, Y) ],
\end{eqnarray*}
which after differentiation with respect to $Y$ yields
\begin{eqnarray}\label{dlla}
  \frac{\partial N({\bf x}_{01},{\bf b}_0, Y)}{\partial Y} =
  \frac{\alpha C_F}{\pi} \,x_{01}^2
  \int_{x_{01}^2}^{1/\Lambda^2_{QCD}} \frac{d x_{02}^2}{(x_{02}^2)^2}
  \, [ 2 \, N({\bf x}_{02},{\bf b}_0, Y) - N({\bf x}_{02},{\bf b}_0,
  Y) \, N({\bf x}_{02},{\bf b}_0, Y) ] ,
\end{eqnarray}
where, for simplicity, we suppressed the difference in the impact
parameter dependence of $N$ on the left and right hand sides of Eq.
(\ref{dlla}). This is done in the spirit of the large cylindrical
nucleus approximation.  Also one should keep in mind that for this
double logarithmic limit in the definition of $N({\bf x}_{01},{\bf
  b}_0, Y)$ given by Eq.  (\ref{NN}) the integration over the dipole's
transverse sizes should be also done from $x_{01}^2$ to
$1/\Lambda^2_{QCD}$.

Now we have to make a connection between $N({\bf x}_{01},{\bf b}_0,
Y)$ and the gluon distribution function $xG_A (x, Q^2)$ of a nucleus.
$N({\bf x}_{01},{\bf b}_0, Y)$ is a forward scattering amplitude of a
$q \overline{q}$ pair on a nucleus and is a well--defined physical
quantity. However there is some freedom in the definition of the gluon
distribution. If one makes us of the general definition of the gluon
distribution as a matrix element of leading twist operator, then an
attempt to take into account higher twist operators would lead only to
renormalization of their matrix elements (see \cite{LRS} and
references therein). The evolution equation for $xG$ would be linear,
with all the non-linear saturation effects included in the initial
conditions. The goal of the GLR type of approach is to put these
non-linear effects in the evolution equation. Therefore in the double
logarithmic approach one usually defines the gluon distribution
function through a cutoff operator product expansion, i.e., as a
matrix element of the $A_\mu A_\mu$ operator, with $Q^2$ an
ultraviolet cutoff imposed on the operator (see the discussion on pp.
442-443 of \cite{MQ}). In the spirit of this approach we define the
gluon distribution by
\begin{equation}\label{xg}
  N({\bf x}_{01},{\bf b}_0, Y) = \frac{\alpha \pi^2}{2 N_c S_\perp}
  {\bf x}_{01}^2 xG_A (x, 1/x^2_{01}),
\end{equation}
with the coefficient fixed by the two gluon exchange between the
quark--antiquark pair and the nucleus (in the large $N_c$ limit).
Substituting Eq. (\ref{xg}) into Eq. (\ref{dlla}) one obtains
\begin{eqnarray*}
  \frac{\partial xG_A (x, 1/x^2_{01})}{\partial Y} = \frac{\alpha
    C_F}{\pi} \, \int_{x_{01}^2}^{1/\Lambda^2_{QCD}} \frac{d
    x_{02}^2}{x_{02}^2} \, [ 2 \, xG_A (x, 1/x^2_{02}) - x_{02}^2
  \frac{\alpha \pi^2}{2 N_c S_\perp} [xG_A (x, 1/x^2_{02})]^2 ].
\end{eqnarray*}
Differentiating the resulting equation with respect to $\ln
(1/x^2_{01}\Lambda^2_{QCD} )$ and using $x_{01} \sim 2/Q$, which is
valid in the double logarithmic limit, we end up with

\begin{eqnarray}\label{GLR1}
  \frac{\partial^2 xG_A (x, Q^2) }{\partial \ln (1/x)\, \partial \ln
    (Q^2 / \Lambda^2_{QCD}) } = \frac{\alpha N_c}{\pi} xG_A (x, Q^2) -
  \frac{\alpha^2 \pi}{S_\perp} \frac{1}{Q^2} [xG_A (x, Q^2)]^2,
\end{eqnarray}
which exactly corresponds to the GLR equation \cite{GLR,MQ}, with the
factors matching those corresponding to cylindrical nucleus case in
references \cite{Levin1,Levin2}.

One has to note that the problems with the definition of the gluon
distribution function outlined above bear no consequence on Eq.
(\ref{eqN}). This equation describes the evolution of $N({\bf
  x}_{01},{\bf b}_0, Y)$ in the leading $\ln (1/x)$ and does not
assume collinear factorization or impose transverse momentum cutoffs,
therefore posing no problems like the mixing of operators of different
twists \cite{LRS}.

One would also like to rederive the equation for $xG_A$ derived
earlier in \cite{Levin1,Levin2,Larry1}, given by Eq. (\ref{glamu1})
here. In the double logarithmic limit we did not reproduce those
results, given by formula (48) in \cite{Levin2}, formula (3.19) in
\cite{Larry1}, and, most explicitly by formula (41) in \cite{JKLW}.
Nevertheless this equation could be obtained from Eq. (\ref{eqN}) in
the following peculiar limit.  Consider the scattering of a virtual
photon with moderately large virtuality $Q$ on a nucleus composed of
very heavy quarks, so that the typical momentum scale characterizing
the nucleus given by the mass of the heavy quarks $M$ is much larger
than $Q$. In that case the double logarithmic limit would correspond
to production of small dipoles in the dipole wave function. Now, using
certain freedom we have in the definition of the gluon distribution
function, we can treat Eq.  (\ref{gamma}b) as a definition of $xG_A$,
which should be substituted in (\ref{gamma}b) instead of $A xG$.
Plugging the generalized Eq. (\ref{gamma}b) into Eq. (\ref{eqN}) taken
in the small--dipole limit we obtain Eq.(\ref{glamu1}).  That way we
will rederive the results of \cite{Levin1,Levin2,Larry1}, however not
quite for the same process as the one for which they were derived
originally.

Finally we note that we failed to find a limit in which Eq.
(\ref{eqN}) reproduces formula (39) in \cite{JKLW},that was given by
Eq.  (\ref{alik}) above, which is another candidate for the double
logarithmic evolution of $xG$ including all multiple pomeron
exchanges.

\section{Conclusions}

\begin{figure}
\begin{center}
\epsfxsize=16cm
\leavevmode
\hbox{ \epsffile{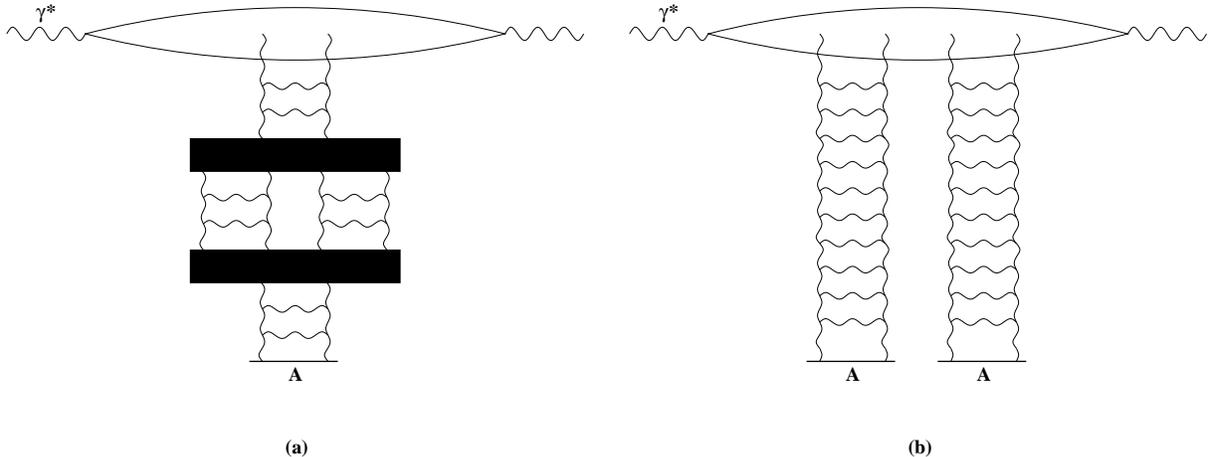}}
\end{center}
\caption{ (a) Diagrams which are not included in our analysis. 
(b) A diagram which is included in Eq. (\ref{eqN}).}
\label{loop}
\end{figure}

One of the diagrams that were not included in our analysis above is
shown in Fig.  \ref{loop} (a). There a hard pomeron ladder splits into
two and then the two ladders again fuse into one which connects to the
nucleus. In general this diagram is of the same order in coupling and
rapidity as the usual two--pomeron exchange diagram of Fig. \ref{loop}
(b), and, therefore should be also considered. The effects of that
type can play an important role in unitarization and saturation of the
onium--onium scattering amplitude \cite{KMW}. In the dipole wave
function language the graph in Fig. \ref{loop} (a) can be interpreted
in two ways. In one case it could be a part of the dipole wave
function, which should be redefined to include this graph in the form
of color quadrupoles. On the other hand the graph in Fig. \ref{loop}
(a) could correspond to the usual ``two dipole'' part of wave function
($n_2$) with both dipoles interacting with the same nucleon in the
nucleus.  However, for the given problem of scattering on a large
nucleus, the graph in Fig.  \ref{loop} (a) is suppressed by powers of
atomic number $A$ compared to the graph in Fig. \ref{loop} (b) for
both cases.

The inclusion of the pomeron fusion effects in the dipole wave
function is a difficult task \cite{KMW}.  Eq.  (\ref{ZZ}) does not
take them into account. Construction of dipole wave function which
includes the diagrams shown in Fig. \ref{loop} (a) is an interesting
and important problem, which is still to be solved. One should note
that inclusion of the graphs of the type shown in Fig.  \ref{loop} (a)
in the dipole wave function would result in appearance of color
quadrupoles \cite{KMW}. In the case of onium--onium center of mass
scattering this wave function can be either considered as suppressed
by factors of $N_C^2$ compared to the one--dipole wave function, or,
more correctly, one may note that its contribution to the scattering
amplitude will suppressed by the factor of $e^{(\alpha_P - 1) Y/2}$
\cite{Mueller3}, with $\alpha_P$ the intercept of the BFKL pomeron.
However, this argument does not apply for the case when one of the
onia is at rest \cite{KMW}, or to our case when the nucleus is at
rest.

The graph in Fig. \ref{loop} (a) in our case of nucleus at rest brings
in suppression by powers of $A$.  The reason for that is very simple:
in the first diagram (Fig. \ref{loop} (a)) there is only one pomeron
ladder (dipole) interacting with the nucleus below, whereas in the
second diagram (Fig. \ref{loop} (b)) there are two ladders. Since each
dipole is convoluted with its propagator through the nucleus, each
ladder brings in a factor of $\gamma ({\bf x} , {\bf b}) d^2 b$.  For
large transverse size dipoles this factor is proportional to $A^{2/3}$
and for small dipoles it scales as $A^1$.  In any case one can see
that the graph in Fig.  \ref{loop} (a), having the same parametrical
dependence on $\alpha$ and $Y$ as the graph in Fig.  \ref{loop} (b),
is suppressed by some power of $A$ compared to this second graph.
Therefore, one should note that by neglecting the effects of the graph
in Fig. \ref{loop} (a) we assume that we are doing the leading
calculation in the powers of the atomic number of the nucleus $A$.
That allows us to avoid complications associated with the
incorporation of the diagram in Fig. \ref{eqN} (a) in the evolution
equation (\ref{eqN}).

Eq. (\ref{eqN}) can also be derived directly from Eq. (\ref{ZZ}) by
putting $u ({\bf x}_{01},{\bf b}_0) = \gamma({\bf x}_{01},{\bf b}_0) +
1$ in it and defining $N = 1 - Z$. That way we have a method of
resumming all multiple pomeron exchanges for any higher order
corrections to the dipole kernel. If one calculates the dipole kernel,
say, at the next-to-lowest order, then we can write down an equation
for generating functional $Z$, similar to Eq. (\ref{ZZ}). Though the
next-to-lowest order equation will in addition have cubic terms in $Z$
on the right hand side. Then, putting $u = \gamma + 1$ and $Z = 1 - N$
one would easily obtain an equation resumming multiple pomeron
exchanges in the subleading logarithmic approximation. Therefore
dipole model provides us with a relatively straightforward way of
taking into account the multiple pomeron exchanges once the
one--pomeron exchange contribution has been calculated. In other words
if the dipole kernel is known at any order in the coupling constant
one can easily generalize the resulting equation for generating
functional to include the multiple pomeron exchanges on a nucleus.

Finally we note that it would be interesting to try fitting the recent
HERA data \cite{Cald} using the evolution of the $F_2$ structure
function given by Eq. (\ref{eqN}).

\section*{Acknowledgments}

I would like to thank Alex Kovner, Genya Levin, Larry McLerran and
Alfred Mueller for a number of informative and very helpful
discussions.  This work is supported by DOE grant DE-FG02-87ER40328.

\end{document}